\documentclass[preprint,12pt]{elsarticle}

\usepackage{graphicx}
\usepackage{amssymb}
\usepackage{lineno}
\usepackage[section]{placeins}

\usepackage[font=footnotesize]{caption}
\usepackage[normalem]{ulem}
\usepackage{color}
\usepackage{amsmath}
\makeatletter
\def\ps@pprintTitle{%
  \let\@oddhead\@empty
  \let\@evenhead\@empty
  \def\@oddfoot{\reset@font\hfil\thepage\hfil}
  \let\@evenfoot\@oddfoot
}
\makeatother

\begin{document}

\begin{frontmatter}


 %\tnotetext[label1]{}
 %\author{Name\corref{cor1}\fnref{label2}}
 %\ead{email address}
 %\fntext[label2]{}
 %\cortext[cor1]{}
 %\address{Address\fnref{label2}}
 %\fntext[label3]{}

\title{A search for neutrino-antineutrino mass inequality by means of sterile neutrino oscillometry}

\author[label1,label2]{\bf M.V. Smirnov}
%\author[label5]{E.Kh. Akhmedov}
\author[label1]{K.K. Loo}
\author[label2,label3]{Yu.N. Novikov}
\author[label1]{W.H. Trzaska}
\author[label4]{M. Wurm}
\address[label1]{\it Department of Physics, University of Jyv\"{a}skyl\"{a}, 40014 Jyv\"{a}skyl\"{a}, Finland}
\address[label2]{St.Petersburg State University, 198504 Peterhof, St.Petersburg, Russia}
\address[label3]{ Petersburg Nuclear Physics Institute, 188300 Gatchina, St.Petersburg, Russia}
\address[label4]{Johannes Gutenberg University Mainz, 55099 Mainz, Germany}

\begin{abstract}
The investigation of the oscillation pattern induced by the sterile neutrinos might determine the oscillation parameters, and at the same time, allow to probe CPT  symmetry in the leptonic sector through neutrino-antineutrino mass inequality.  We propose to use a large scintillation detector like JUNO or LENA to detect electron neutrinos and electron antineutrinos from MCi electron capture or beta decay sources. 
Our calculations indicate that such an experiment is realistic and could be performed in parallel to the current research plans for JUNO and RENO. Requiring at least 5$\sigma$ confidence level and assuming the values of the oscillation parameters indicated by the current global fit, we would be able to detect  neutrino-antineutrino mass inequality  of the order of 0.5\% or larger,  which would imply a signal of CPT anomalies.
\end{abstract}


\end{frontmatter}

%%\linenumbers

\section{Introduction}
\label{S:1}

The discovery of neutrino oscillations \cite{sup_kam} provided undisputed evidence of incompleteness of the conventional Standard Model (SM) of particle physics \cite{SM} and gave a clear indication that further neutrino studies may lead us towards phenomena Beyond the Standard Model (BSM). One of such possible manifestations would be violation of CPT invariance \cite{Kostelecky_cpt} in the process of neutrino oscillations.

CPT symmetry is considered a fundamental law of nature, to which we know no exceptions yet. One of the important implications of this symmetry
law is the requirement that in all processes involving particles and their antiparticles the corresponding probabilities and other parameters governing the process should be exactly the same. Many experiments based on that principle have been conducted to probe the CPT conservation. For instance, in the leptonic sector, a measurement of \textit{g}-factor for electron and positron performed at the University of Washington in Seattle \cite{g-factor_exp} or, in the baryonic sector, a mass measurement of proton and antiproton implemented at CERN \cite{proton_antiproton,blaum}. 

Neutrinos carry important information about weak interactions. Especially their interferometric nature, make them a very sensitive tool to search for unconventional physics, in particular CPT violation and Lorentz violation \cite{Kostelecky_cpt,Greenberg}.
%Nevertheless the issue of CPT violation in the leptonic sector remains unresolved.
However, until now there was no direct and simultaneous experiment with comparison of neutrino and antineutrino, therefore the issue of CPT violation in the leptonic sector remains unresolved. The mass inequality could give a unique possibility to do it.

Obviously, the properties of neutrinos doom the attempts to verify CPT conservation through oscillation measurements to a likely failure or, at best, have a prohibitively high price tag. Very low interaction cross-sections, 
limited neutrino flux, and the size of the available detectors exclude such attempts on short time scales. However, if the active neutrinos like the well-known electron-, muon-, and tau neutrino and their antiparticles were not the only existing neutrino flavours, it might be possible to search for CPT violation already in the near future, as it is outlined in this article.

One of the exciting new hypotheses is the existence of the light sterile neutrinos with unknown nature and properties \cite{wp_st_neut}. 
The existence of one or more sterile flavors is indicated by the re-evaluation of reactor neutrino fluxes experiments \cite{st_anom}, some of the beam experiments such as MiniBoone, LSND  and  the Gallium anomaly \cite{wp_st_neut,lsnd_miniboone}.

Although the direct observation of sterile neutrinos is not possible, their presence would be reviled by a distinct oscillation pattern, caused by the conversion from active to sterile flavor. If observed, the pattern would also give a hint of the number of sterile neutrino flavors. The simpliest assumption is the scenario with one additional sterile flavour, so called 3+1 scheme \cite{3+1}.  The current global fit to the oscillation data suggests that the additional mass-squared difference and mixing angle governing the active-to-sterile conversion should be of the order of
$\Delta m_{41}^2$ $\sim 1~\textrm{eV}^2$ and $\sin^2 2\theta_{ee}$ $\sim 0.1$ respectively \cite{Kopp:2013vaa}.  Assuming these parameter values and relying on the proportionality between the energy and oscillation length, we may choose the energy of neutrinos in a way that at least one full oscillation period will overlap with the active volume of a single detector, like it is proposed in the Borexino SOX experiment \cite{sox}.    The realization of this neutrino oscillometry method  \cite{neu_oscil} will open a unique opportunity to search not only for disapperance oscillations to sterile neutrinos but also to check the validity of CPT conservation in the leptonic sector if such a transition is observed.

\section{Probing neutrino-antineutrino mass inequality}
\label{S:2}

If CPT is strictly conserved the oscillation of neutrinos and antineutrinos 
would be described by identical probability functions with two main parameters: mixing angles and mass-squared differences. For instance, in the case of 3+1 scenario, the survival probability of electron neutrino and electron antineutrino can be approximated with  \cite{neu_oscil}:
\begin{equation}
\label{osc_neutr}
P(\overset{(\--)}{\nu}_e \to\overset{(\--)}{\nu}_e)=1-\sin^2(2\theta_{ee}) \cdot\sin^2(1.27\cdot\Delta m_{41}^2\cdot \frac { L}{ E}).
\end{equation}
Note that the equation (\ref{osc_neutr}) is valid for short baseline experiment (SBL). 
Consequently, if CPT conservation holds, we should obtain in the classical case \cite{Giunti_cpt}: 
\begin{equation}
\label{ratio}
\frac{( \Delta m_{41}^2)_N}{(\Delta m_{41}^2)_A}=1, 
\end{equation}
where $N$ stands for the value obtained from the measurement with neutrinos and $A$, with antineutrinos.  
At the same time, if in the real experiment the mass inequality will be observed, then it would mean CPT violation and the estimation of this process
should be done within the framework of the Standard Model Extension (SME) formalism with using coefficients of Lorentz- and CPT-violation \cite{Kostelecky_cpt}.
In this paper, the comparison is  made only for the ratio of the mass-squared differences, because
$\Delta m_{41}^2$ can be extracted from the experimental data with better precision than $\Theta_{ee}$. For instance, it depends far less on the intensity normalization.  
The evaluation of $\Theta_{ee}$ will be considerably less accurate, because of the systematic uncertainties in absolute values of neutrino flux and fiducial volume.

It should be noted, that $P(\nu_e\to\nu_e)\ne P(\bar\nu_e\to\bar\nu_e)$ 
would mean both CPT and CP violation. 
However the latter can not be probed in survival experiments if CPT is conserved.\footnote[1]{suggested by E.Kh. Akhmedov}

\section{Outline of the experiment}
\label{S:3}

Successful application of neutrino oscillometry requires some degree of energy selection and well-defined source position. For these reasons the experiment cannot rely on natural neutrino sources such as the Sun, the Earth or cosmic neutrinos.  Nuclear power reactors also are excluded, as they do not produce considerable amounts of electron neutrinos. The best alternatives are provided by man-made, high-intensity beta decaying radioactive sources emitting electron neutrinos and antineutrinos with the energy around 1 MeV. This choice is based on the outcome of the global fit predicting $\Delta m_{41}^2$ of the order of 1 $\textrm{eV}^2$. Precise measurement of that value both for electron neutrinos and antineutrinos will be the key outcome of the proposed experiment.

The mass-squared difference will be derived from the reconstructed oscillation probability curves. To measure the oscillation pattern the detector has to have sufficiently low energy threshold for the registration of both neutrinos and antineutrinos and have adequate position and energy resolution.  Furthermore, to achieve the required statistics, the active volume of the detector should ideally extend over at least one oscillation period. The only devices that can currently fulfil these requirements are large liquid scintillator detectors.

\subsection{Neutrino and antineutrino sources}
\label{S:3_1}
\begin{figure}[ht]
\centering
\begin{minipage}[b]{0.45\linewidth}
\centering
\includegraphics[scale=0.59]{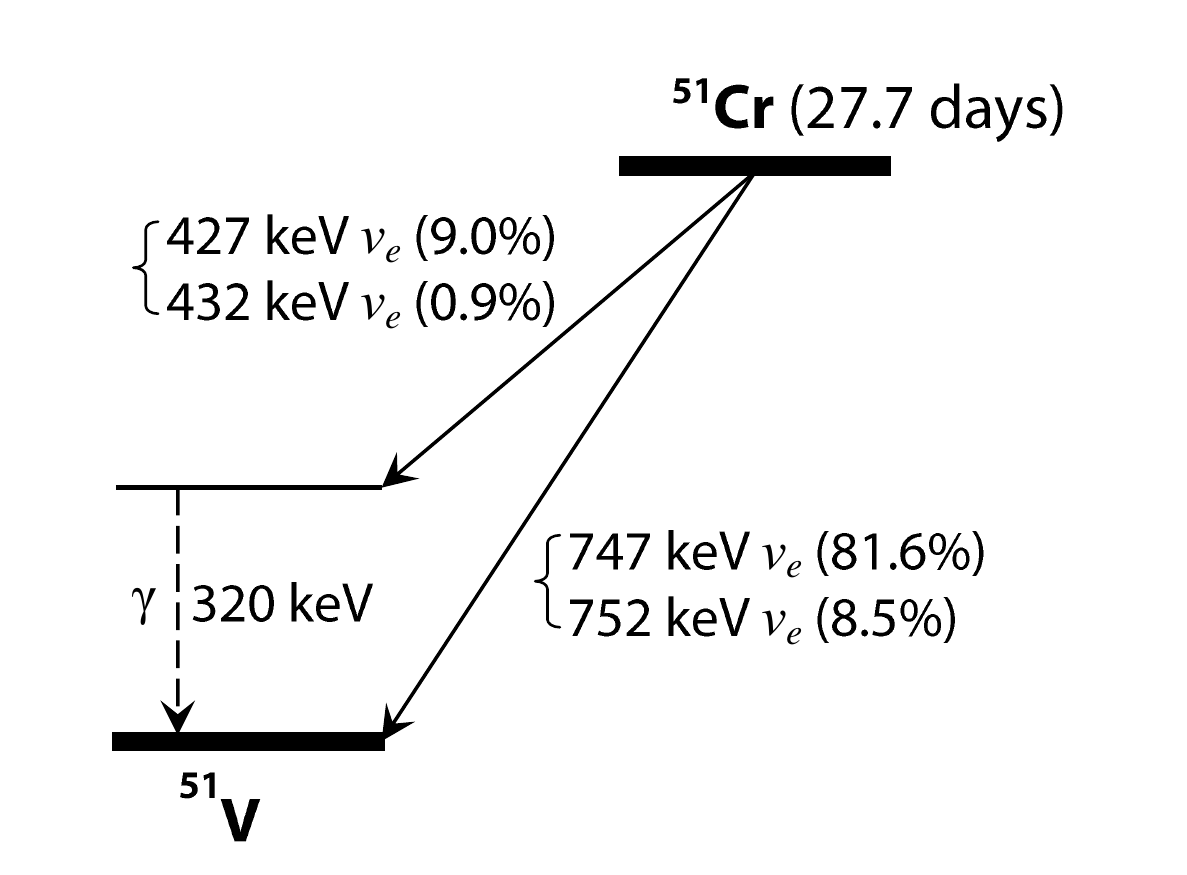}
\caption{Figure 1. The decay scheme of $\rm{^{51}Cr}$, with indicated energies of emitted neutrinos \cite{51cr}.}
\label{fig:cr_decay}
\end{minipage}
\hfill
\begin{minipage}[b]{0.45\linewidth}
\centering
\includegraphics[scale=0.45]{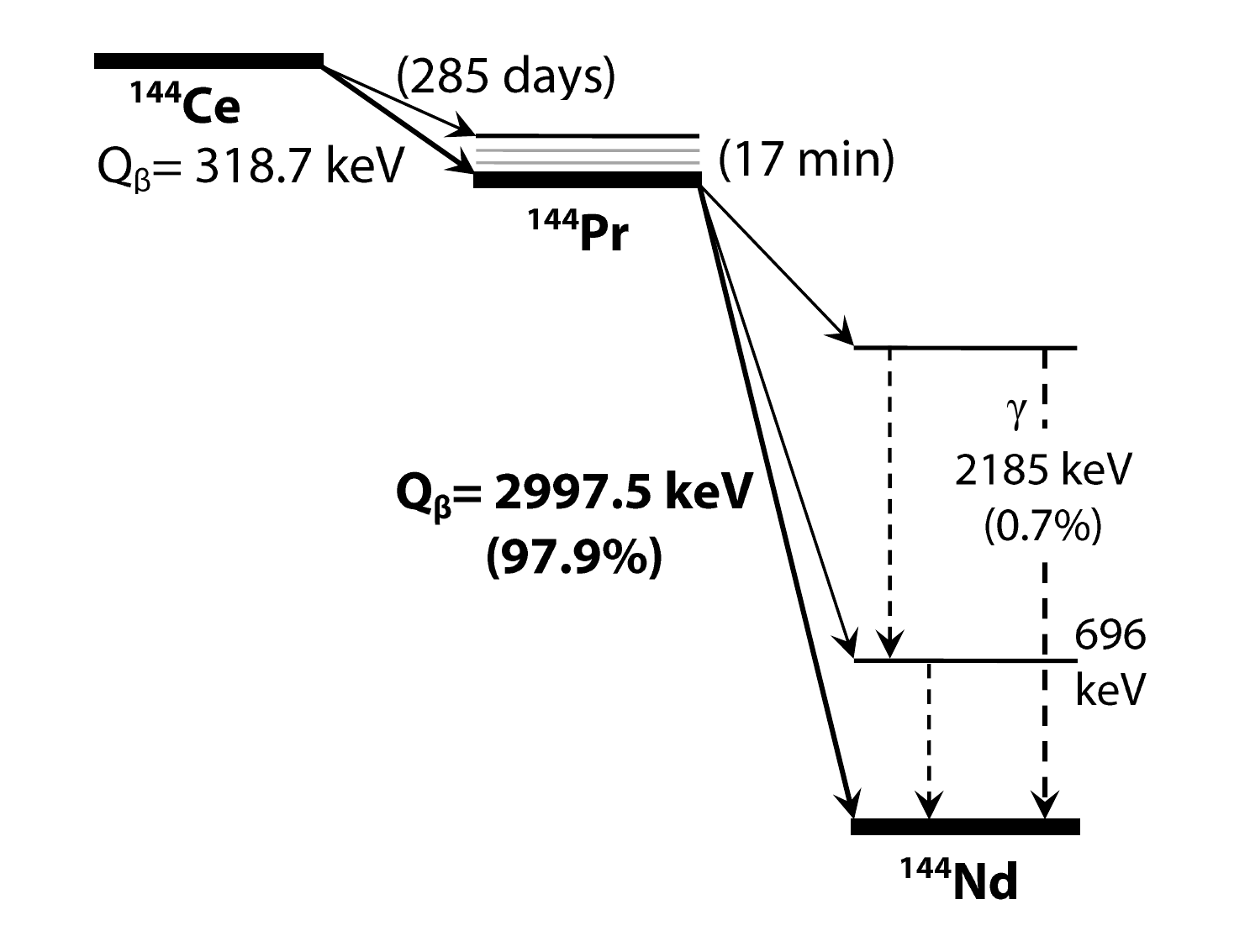}
\caption{Figure 2. The decay scheme of system $\rm{^{144}Ce-^{144}Pr}$ \cite{Pr_data,Ce_source}.}
\label{fig:ce_decay}
\end{minipage}
\end{figure}

The obvious first choice for a $\nu_e$ source is $\rm{^{51}Cr}$, well known from GALLEX \cite{gallex,gall_51cr} and GNO \cite{gno} experiments. The decay scheme of $\rm{^{51}Cr}$ is shown in Fig.~\ref{fig:cr_decay}. This nuclide has 27.7 days half-life and a neutrino spectrum dominated by mono-energetic lines at an energy of around  $\sim$0.75 MeV (90.1\% total branching ratio), as shown in Fig.~\ref{fig:cr_decay}. Current estimations for the SOX project\cite{sox} indicate that a 35 kg chromium oxide source enriched in $^{50}$Cr ($\sim$30\%) placed in a nuclear reactor, can be activated up to 10 MCi. 
Accounting for the decay during extraction from the reactor, transport, and installation inside the detector, the realistic activity of the source at the start of the experiment would
be around 8 MCi.

As the source of $\bar\nu_e$ we propose $\rm{^{144}Ce-^{144}Pr}$ chain \cite{Ce_source}.  
The decay scheme is shown in Fig.~\ref{fig:ce_decay}. The antineutrino energy spectrum of $\rm{^{144}Pr}$ is continuous with the end point energy at about 3 MeV \cite{spect} and with the overall half-life of 285 days.  
About 48.5\% of the emitted antineutrinos are at energies above the detection threshold for the IBD reaction (see Section \ref{S:3_2}) and can be used in the experiment. Current studies indicate that an activity of 100 kCi or more can be reached for the Ce-Pr source \cite{Ce_new_paper}. In our calculations we assume the initial activity of 120 kCi.

The properties of both sources are summarized in Table~\ref{tabular:prop}.
\renewcommand{\arraystretch}{1.2}
\begin{table}[h]
\begin{center}
\caption{Properties of sources proposed  as neutrino-  and antineutrino emitters for the oscillometry measurement.}
\label{tabular:prop}
\begin{tabular}{|c|c|c|c|c|c|c|}
\hline
Source & Type & Spectrum & E, MeV & A, MCi & $\rm T_{1/2}$, d  & Expos., d\\
\hline
$\rm{^{51}Cr}$ & $\nu_e$ & mono & 0.75 & 8 & 27.7 & 55 \\
\hline
$\rm{^{144}Ce-^{144}Pr}$ & $\bar\nu_e$ & continuous & 1.8 - 3.0 & 0.12 & 285 & 300 \\
\hline
\end{tabular}
\end{center}
\end{table}

\subsection{Neutrino detection in a liquid scintillation detector}
\label{S:3_2}

Elastic neutrino-electron scattering is the dominant interaction channel leading to the registration of $\nu_e$ events in a liquid scintillator detector. This reaction
$$
\nu_e + e^- \rightarrow \nu_e +e^-
$$
has similar characteristic to the Compton effect. While all three neutrino flavours  interact with electrons by neutral current, electron  neutrinos interact as well via charged current. The cross-section of this interaction may be approximated by a simple expression: $\rm\sigma=0.7\times E_\nu\cdot10^{-44}cm^2$ \cite{elastic_scatter_form}.
The uncertainty of this estimation, including of radiation corrections, is about 2\%  \cite{elastic_scatter_form}.

The dominant detection channel of electron antineutrinos in liquid scintillator is the inverse beta decay reaction  (IBD)
$$
\bar\nu_e + p \rightarrow n+e^+
$$
This reaction produces two distinct signals in the detector allowing for an unambiguous identification of a neutrino event. The kinetic energy and annihilation of the positron and the second, from neutron capture on hydrogen. 
The former is prompt and contains the information on the energy of the neutrino. The later 2.2 MeV gamma signal is delayed by about 250 $\rm\mu s$.
The threshold  energy for IBD is 1.8 MeV and  the cross section  of the  process  can  be approximated by the expression
$\rm\sigma=9.5\times(E_\nu-1.29)^2\cdot10^{-44}cm^2$ \cite{IBD_cross}.

\subsection{Proposed implementation and expected event rates}
\label{S:3_3}

To evaluate the feasibility of the proposed experiment the event rates were calculated for two alternative shapes of liquid scintillator detector: spherical, with the source installed in the centre, and cylindrical, with the source mounted close to the center of the end cap of the tank.

\begin{figure}[ht]
\centering
\begin{minipage}[b]{0.45\linewidth}
\center{\includegraphics[scale=0.45]{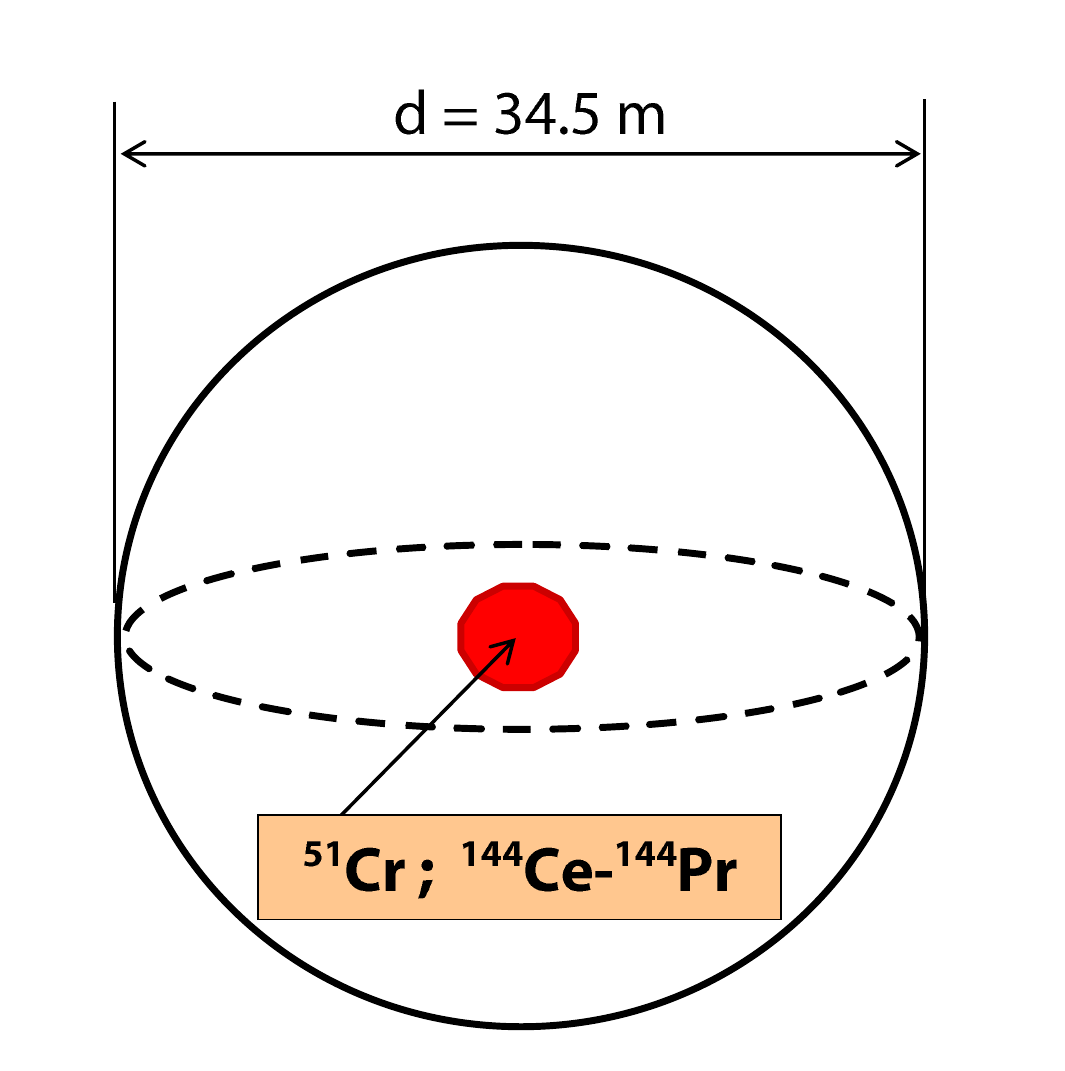} \\ A.}
\end{minipage}
\hfill
\qquad
\begin{minipage}[b]{0.45\linewidth}
\center{\includegraphics[scale=0.45]{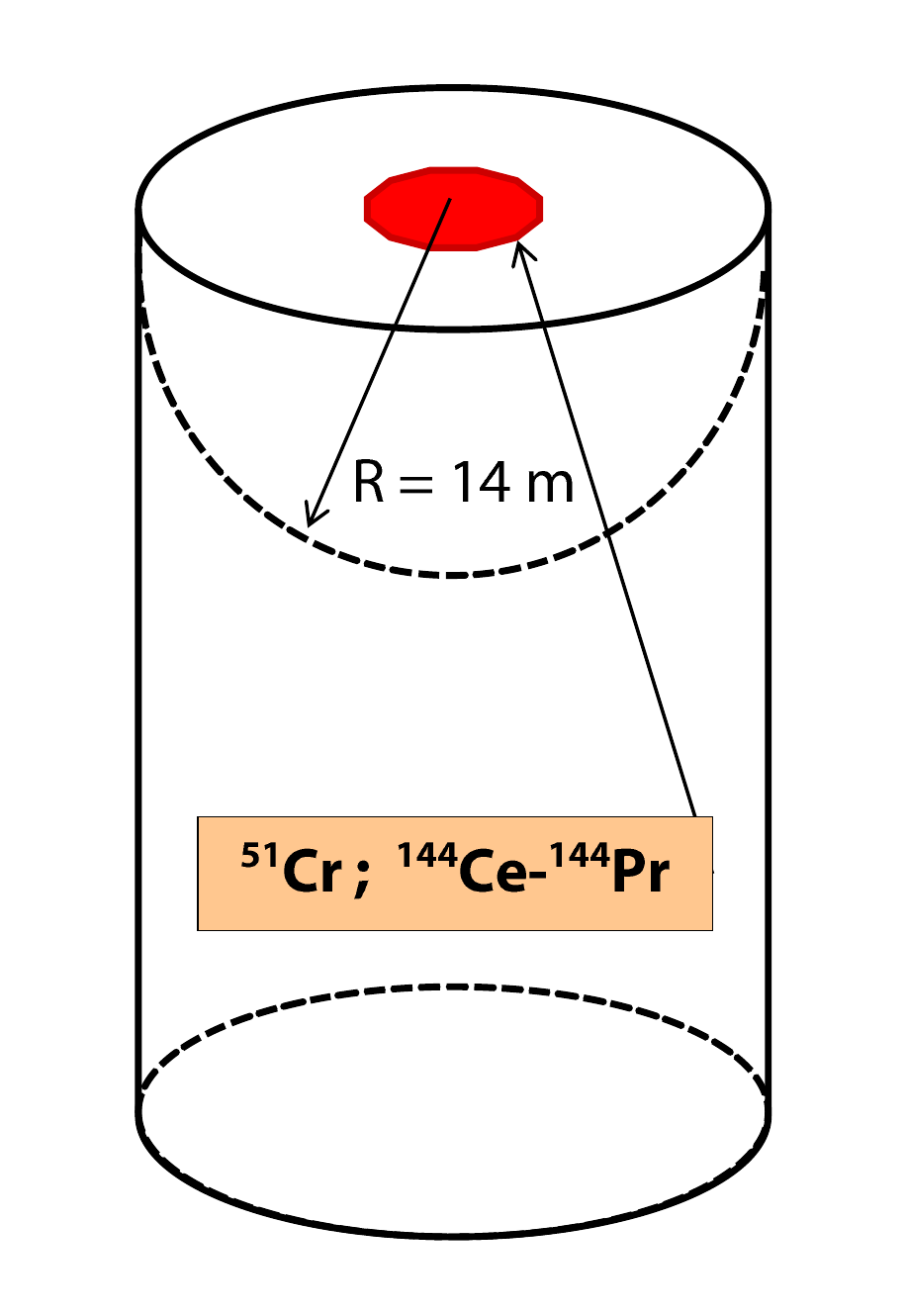} \\ B.}%fig 5

\end{minipage}
\caption{Figure 3. The scheme of the proposed experiment. Left panel: the configuration for JUNO with the source in the center of the detector. Right panel: the configuration for LENA with the source in the center of the top of the tank.}
\label{fig:exp_sch}
\end{figure}

The specifications of the spherical detector are those of JUNO \cite{juno}, currently under construction in China. It will contain 20 ktons (fiducial) of liquid scintillator enclosed in a spherical volume with a diameter of 34.5 m.  
Spherical shape provides the optimal detection geometry when the source of $\nu_e$ or $\bar\nu_e$  is placed in the center, as shown on Fig.~\ref{fig:exp_sch}A. The expected exposition time is restricted by the half-life. For the Cr source we therefore assume 55 days and for the Ce-Pr source 300 days. Gamma-ray  background  from the source inside the detector  can be reduced  by using a radiation shielding, e.g. from tungsten  \cite{sox}.

To evaluate the case of a cylindrical detector, we have used LENA specifications \cite{Wurm}: 100 m high, 28 m diameter, 50 kton fiducial mass. The source of $\nu_e$ or $\bar\nu_e$ would be located close to the top of the detector, as it is shown on Fig.~\ref{fig:exp_sch}B.
%PICTURE
In this case only the hemispherical volume near the top would be considered for analysis. This is clearly a less favourable geometry as it uses only part of the fiducial volume and has to cope with non-optimal coverage by the photo sensors (PMTs). The only considerable advantage of this geometry is that it eliminates the need to immerse the source into the volume of the detector. For the purpose of this analysis the two detector geometries differ just by a factor of 4 (volume of a sphere vs. hemisphere) hence energy- and distance-dependent event rate in both cases can be expressed by the equation \cite{Ce_source,oscil_source}:
\begin{equation}
\label{rate}
N(L, E)=\int\limits_{L-\frac{\Delta}{2}}^{L+\frac{\Delta} {2}}\int\limits_{E_{min}}^{E_{max}}\frac{A_0}{\lambda}\cdot n\cdot\sigma(E)\cdot S(E)\cdot P(L, E)\cdot(1-\exp[-\lambda t_e] )\,dE\,dL,
\end{equation}
where $A_0$ is the source activity at the start of the measurement, \textit{n} the density of free protons (electrons) in the target, $\Delta$ the width of bin with the center $L$, $\sigma(E)$  the cross section for IBD or electron-neutrino scattering, $S(E)$  the spectral shape (delta-function for monoenergetic neutrinos), $P(L, E)$  the oscillation probability (\ref{osc_neutr}), $t_e$  the time of measurement and $\lambda$  the decay constant of the source.

\begin{figure}[h!]
\centering
\begin{minipage}[b]{0.45\linewidth}
\center{\includegraphics[scale=0.36]{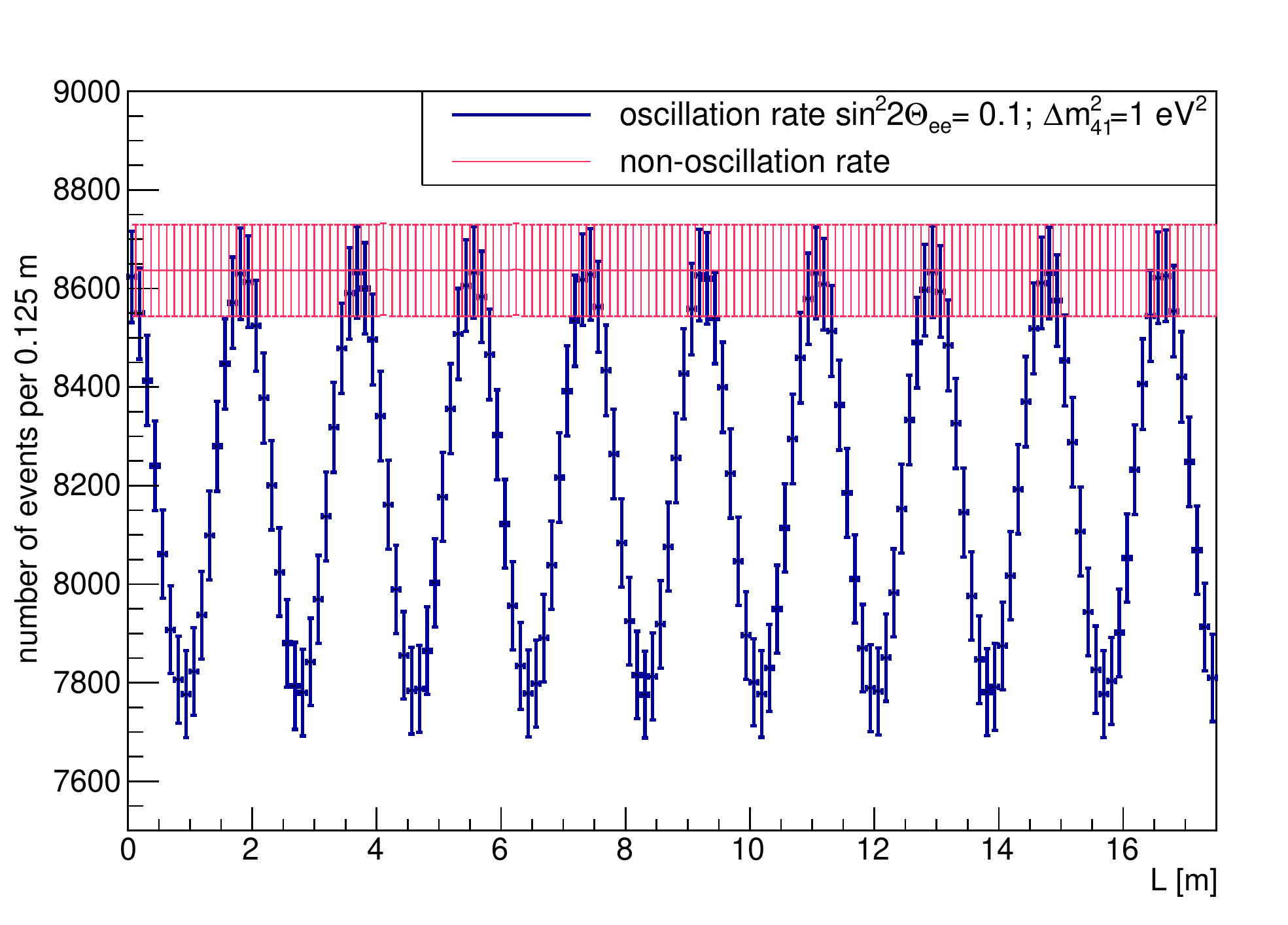}\\ A.}
\end{minipage}
\hfill
\begin{minipage}[b]{0.45\linewidth}
\center{\includegraphics[scale=0.36]{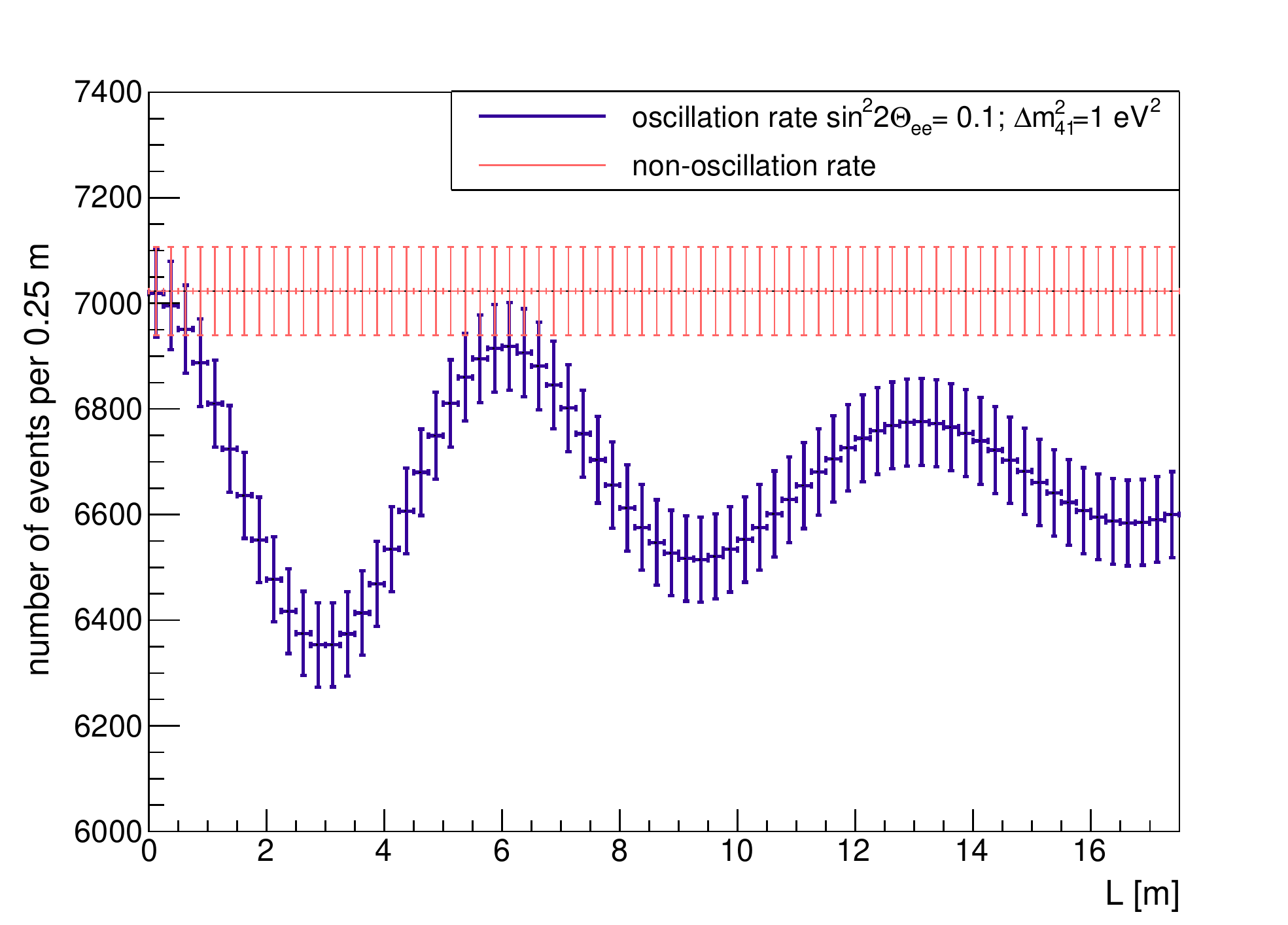}\\ B.}
\end{minipage}
\caption{Figure 4. The expected rate in JUNO detector. Left panel: the total rate for $\rm ^{51}Cr$ source, time of measurement 55 days. Right panel: the total rate for $\rm ^{144}Ce-^{144}Pr$ source, time of measurement 300 days. }
\label{fig:rate_juno}
\end{figure}

The total exposure time needed to complete the measurement would be about 355 days assuming that both sources  are used subsequently.  
The event rates were plotted as a function of the distance from the source and are shown on Fig.~\ref{fig:rate_juno}A and Fig.~\ref{fig:rate_juno}B for $\nu_e$ and $\bar\nu_e$ respectively. The presented calculations correspond to the spherical geometry.
%PICTURE

Since the detection channels for $\nu_e$ and $\bar\nu_e$ are different and disentangled, it would be possible to make the measurement with both radioactive sources at the same time.This configuration would even have the advantage of reduced systematic error due to normalization. However, from the implementation point of view, that would be a major experimental challenge.

\section{Results}
\label{S:4}
The main outcome of the experiment will be the independent determination of the oscillation parameters for neutrinos and antineutrinos by fitting the spectra shown on Fig.~\ref{fig:rate_juno}.
As it was explained earlier, the main parameter for testing CPT symmetry is $\Delta m_{41}^2$.
The confidence level can be evaluated from equation (\ref{ratio}),
\begin{equation}
\label{ratio_error}
\frac{(\displaystyle \Delta m_{41}^2)_N\pm\sigma_N}{(\displaystyle \Delta m_{41}^2)_A\pm\sigma_A}=1\pm\sigma,
\end{equation}
where $\sigma_N$ is the uncertainty of measurement of $(\Delta m_{41}^2)_N$, $\sigma_A$ the uncertainty of measurement of $(\Delta m_{41}^2)_A$, $\sigma$ the total uncertainty of ratio. Since the starting assumption is that $(\Delta m_{41}^2)_N$ and $(\Delta m_{41}^2)_A$ are identical (although with different  error bars), the relative uncertainty can be estimated  as follows:
\begin{equation}
\label{total_error}
\sigma\le\frac{\sigma_N}{(\Delta m_{41}^2)_N}+\frac{\sigma_A}{(\Delta m_{41}^2)_A}.
\end{equation}

The event samples were generated by using Monte-Carlo  approach. The basic steps which were used in analysis:
\begin{itemize}
\item The oscillation spectrum was generated with certain oscillation parameters ($\Delta m_{41}^2$ and $\sin^2 2\theta_{ee}$).
\item The pattern of real oscillation spectrum was obtained with adding of basic uncertainties: statistical error, the error of position resolution, the error of source activity and the error of fiducial volume of detector. For antineutrino case the energy resolution was also used in analysis.
\item The oscillation parameters 
were extracted from the pattern of real spectrum  with using fit-function the same as probability function (\ref{osc_neutr}) taking energy and position resolution smearing into account.
\item  The fit-function was optimized by using the method of minimization $\chi^2$  from package Minuit2 ROOT.
\item The gaussian distribution for uncertainties of oscillation parameters was assumed.
\end{itemize} 
%PICTURE
\begin{figure}[h!]
\centering
\begin{minipage}[b]{0.45\linewidth}
\centering
{\includegraphics[scale=0.36]{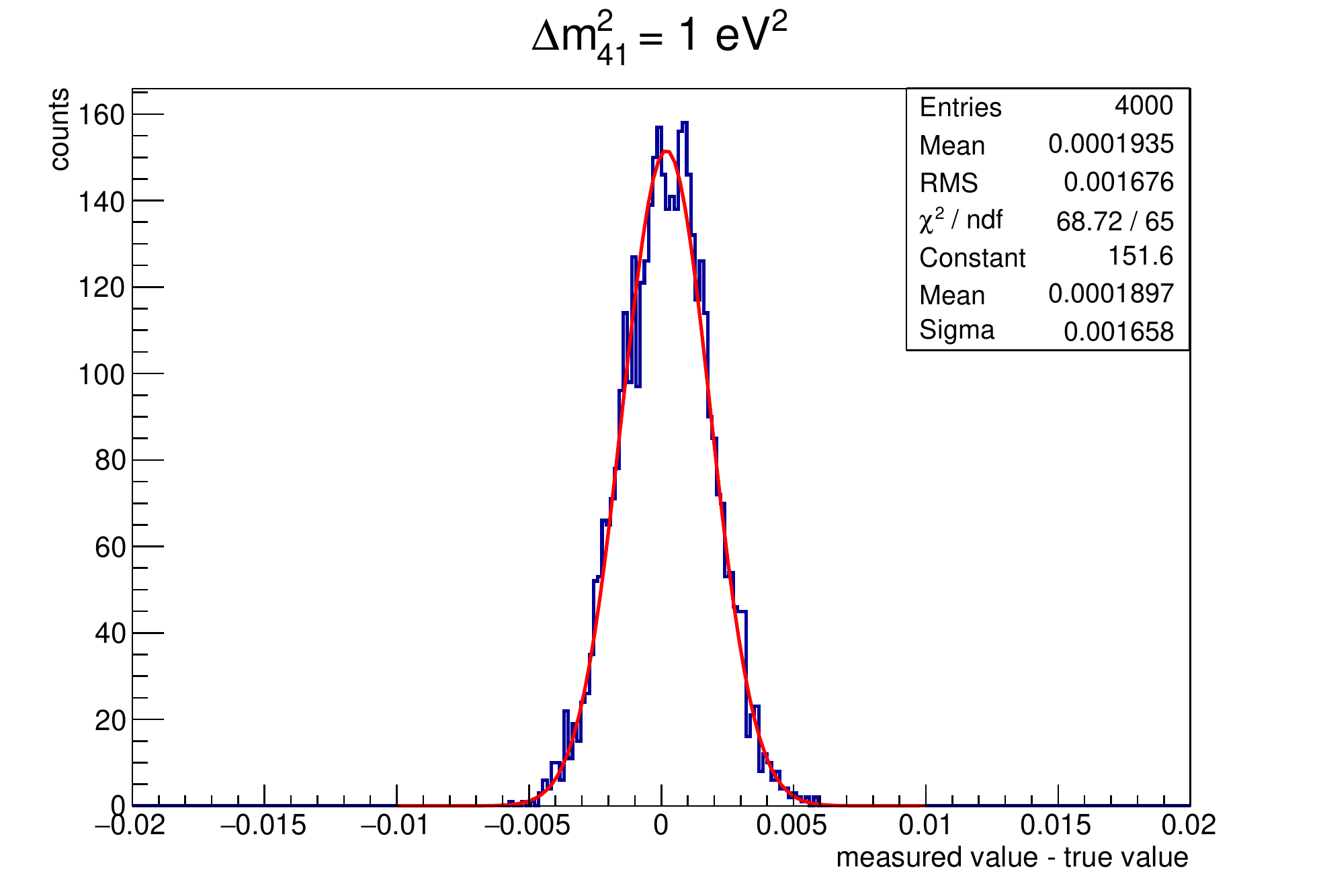}\\ A.}
\end{minipage}
\hfill
\begin{minipage}[b]{0.45\linewidth}
\centering
{\includegraphics[scale=0.36]{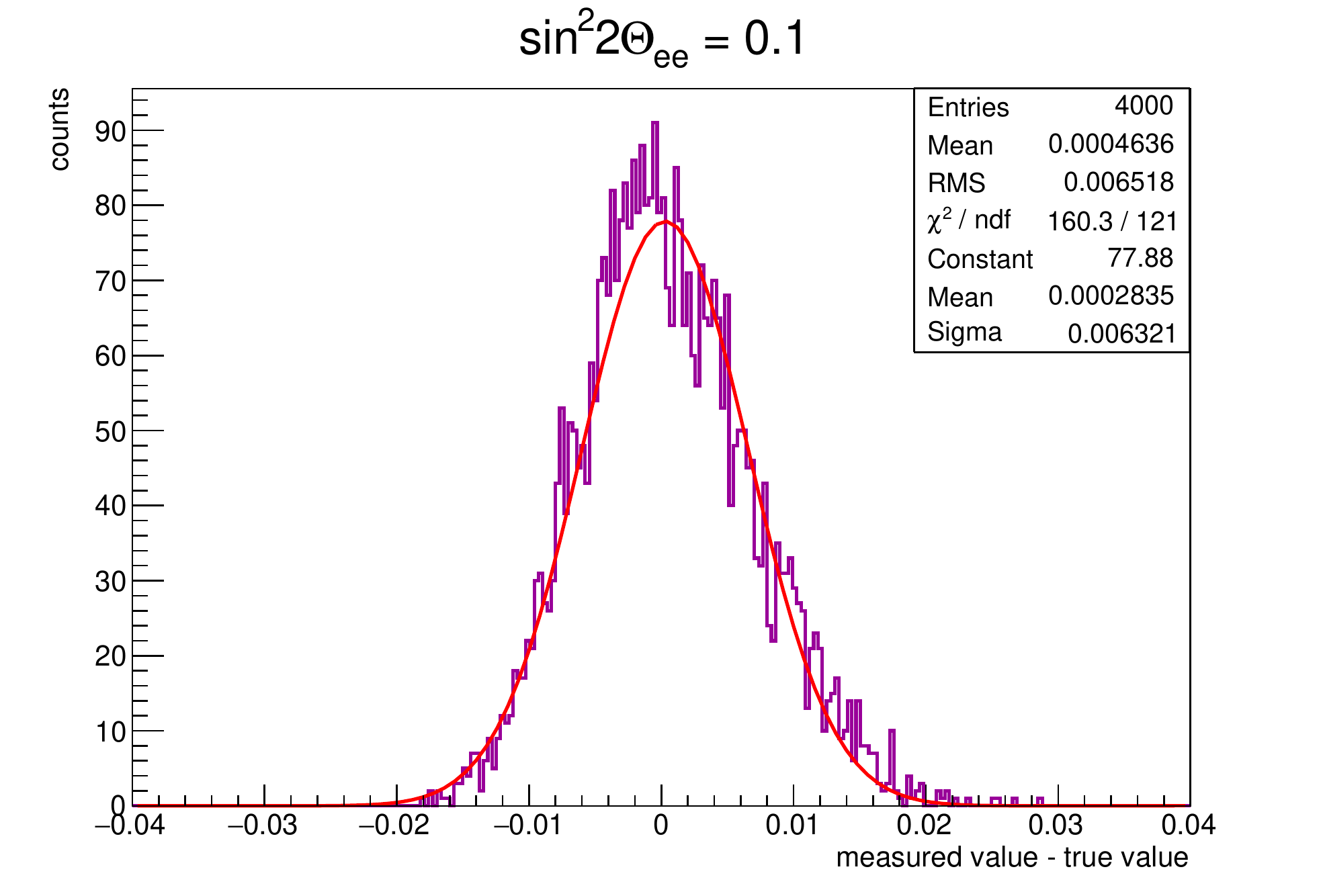}\\ B.}
\end{minipage}
\caption{Figure 5. The distribution of difference between reconstructed value and true value for neutrino case in LENA detector geometry. Left panel: the reconstructed parameter is $\Delta m_{41}^2$ with true value 1 eV$^2$. Right panel: the reconstructed parameter is $\sin^2 2\theta_{ee}$ with true value 0.1.}
\label{fig:gauss}
\end{figure}
The example of such distribution of errors is shown on Fig.~\ref{fig:gauss}. As can be seen, the relative error for $\sin^2 2\theta_{ee}$ is larger at least by one order of magnitude than the relative error for $\Delta m_{41}^2$. It explains also our choice for $\Delta m_{41}^2$ as the key paprameter parameter of CPT test.  The uncertainty  of these measured values of $\Delta m_{41}^2$ was determined independently for both sources, and combined with the equation (\ref{total_error}). However this approach allows also to determine  $\Theta_{ee}$ with high precision.

For detector-related uncertainties, the design parameters given by JUNO and LENA collaborations were used.
In the case of JUNO the energy resolution of ~$3\%/\sqrt{E [MeV]}$ and position resolution of  9 cm($\nu_e$) and 4.5 cm($\bar\nu_e$) were used. For LENA  $6.1\%/\sqrt{E [MeV]}$, 10 cm($\nu_e$) and 4.5 cm($\bar\nu_e$)  were used respectively.

Neglecting background from radioactive decays, the main background in the energy region of interest for $\nu_e$ detection arises from the solar neutrinos ($^7$Be). It is shown in Table~\ref{tabular:backgr}. The influence of this factor on the final result was also taken into account in our estimations. Nevertheless, direct  measurements of the solar background are carried out automatically whenever the sources are removed so the correction may be done afterwards. In the case of antineutrinos, the highest  background  comes from nuclear reactors  and  can  be neglected,  even in the case of JUNO  detector,  because the expected neutrino flux from $\rm{^{144}Ce-^{144}Pr}$ source located in the middle of the detector is considerably higher than that of 12 power reactors 53 km away. The estimated event rates from sources with the strength of 8 MCi for $\rm^{51}Cr$ and 0.12 MCi for $\rm{^{144}Ce-^{144}Pr}$ as well as the corresponding  backgrounds are presented in Table~\ref{tabular:backgr}.

\renewcommand{\arraystretch}{1.1}
\begin{table}[!ht]
\begin{center}
\caption{Expected background and signal events for full time of the experiment. The mass of JUNO's sphere is 20 kt, mass of semisphere in LENA is 5 kt.}
\label{tabular:backgr}
\begin{tabular}{|c|c|c|c|c|c|c|}
\hline
Type & Signal & Background  & Signal & Background & Exposition\\
& JUNO & JUNO &  LENA& LENA & time, d\\
\hline
$\nu_e$ &  1148860& 348032 & 459234 & 87008  & 55\\
\hline
$\bar\nu_e$ & 466800 & 3090 & 193500 & 45 & 300 \\
\hline
\end{tabular}
\end{center}
\end{table}

\begin{figure}[ht]
\centering\includegraphics[scale=0.53]{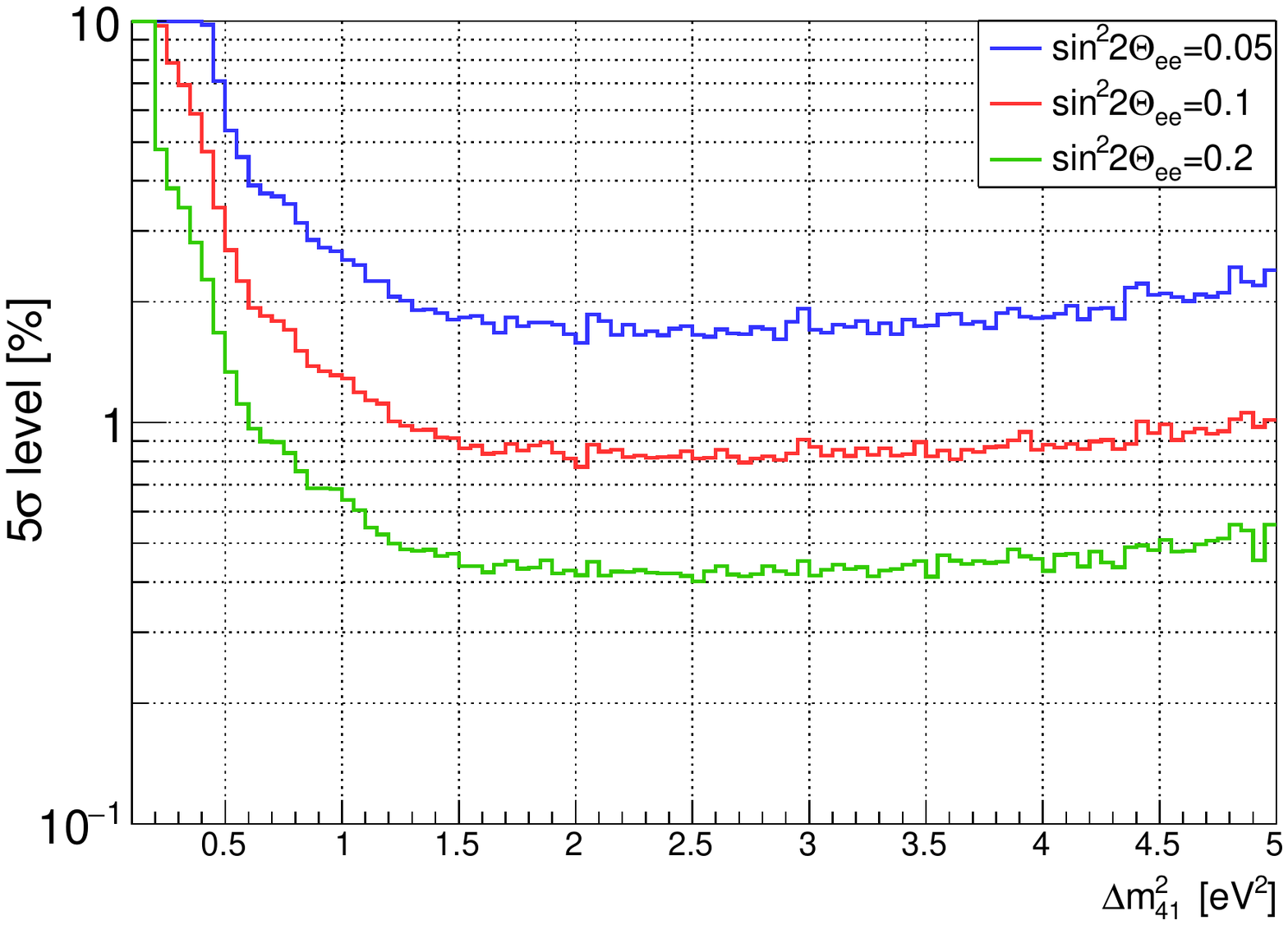}
\caption{Figure 6. Sensitivity limits of LENA for the detection of mass inequality at 5$\sigma$ level of measured events calculated for 3 different mixing angles and plotted as a function of $\Delta m_{41}^2$.}
\label{fig:5s_lena}
\end{figure}

\begin{figure}[hb]
\centering\includegraphics[scale=0.53]{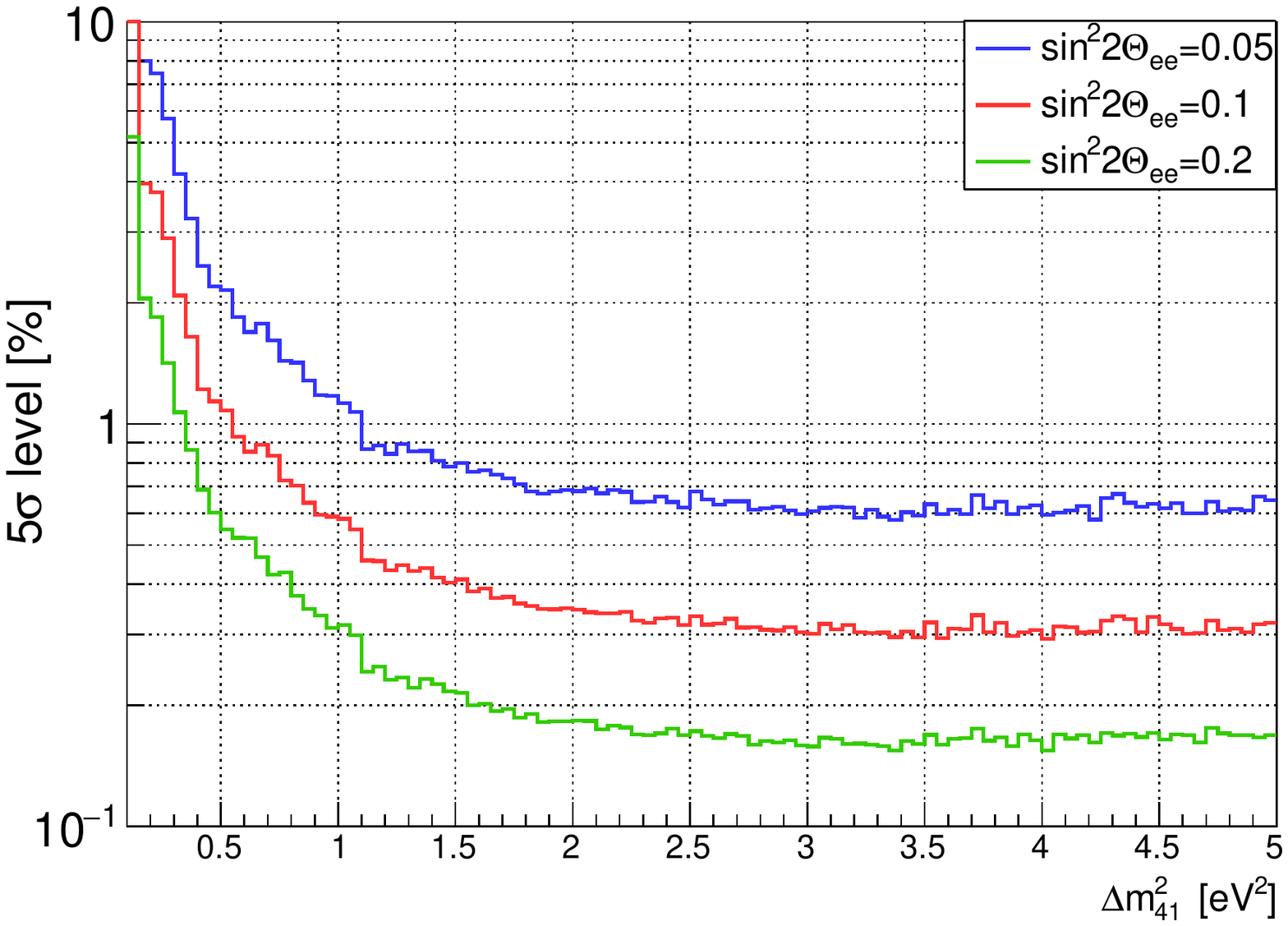}
\caption{Figure 7. Sensitivity limits of JUNO for the detection of mass inequality at 5$\sigma$ level of measured events calculated for 3 different mixing angles and plotted as a function of $\Delta m_{41}^2$.}
\label{fig:5s_juno}
\end{figure}

The main outcome of our simulations is shown in Fig.~\ref{fig:5s_lena}  (for LENA) and Fig.~\ref{fig:5s_juno} (for JUNO). The figures show a 5$\sigma$ sensitivity limit calculated for 3 different $\sin^2 2\theta_{ee}$ values and plotted as a function of $\Delta m_{41}^2$.
For instance, if we check the performance for the parameters chosen by the global fit that is
$\Delta m_{41}^2\ge1 ~\textrm{eV}^2$ and $\sin^2 2\theta_{ee}\approx0.1$, we conclude from Fig.~\ref{fig:5s_lena} that LENA will detect (at the 5$\sigma$ level of measured events) a difference of  $\Delta m_{41}^2$ neutrino/antineutrino and therefore CPT violation that exceeds 1\%. For the same parameters JUNO would reach 0.5\% sensitivity as it may be observed from Fig.~\ref{fig:5s_juno}. The slightly better sensitivity of the JUNO detector is a consequence of the spherical shape and better expected energy resolution.

\section{Conclusions}
 We have outlined a scheme to probe the mass inequality by analyzing the neutrino and antineutrino oscillation patterns induced by the still hypothetical sterile neutrinos. 
 If the "sterile" oscillation patterns are detected, the experiment will yield new fundamentally important oscillation parameters and will be sensitive to the mass inequality on the level of a few tenths of a percent at 5$\sigma$ significance. This sensitivity can be achieved assuming the current best-fit values of the oscillation parameters indicated by the global fit. 

The proposed experiment would utilize a large scintillator like future JUNO, LENA or RENO-50 experiments. Since JUNO is now approved and has just entered the first construction phase, the proposal is certainly feasible. The same concerns the use of MCi sources. Since similar ones were already produced and used by GALLEX and GNO experiments, there is already the necessary knowhow and, in the case of Cr source, also enriched material in sufficient quantity.
\\
\\
\textbf{Acknowledgements}\\
This work was supported by CIMO Fellowship TM-13-8925 grant. M.S. would like to say great thanks to the Department of Physics, University of Jyv\"{a}skyl\"{a} for hospitality. We express our appreciation to E.Kh. Akhmedov for important remarks and corrections and to Ya.I. Azimov, J.S. Diaz for valuable discussions.


\bibliographystyle{model1-num-names}
\bibliography{sample.bib}

\begin{thebibliography}{00}
	\bibitem[1]{sup_kam}
	Y. Fukuda \textit{et al.} (Super-Kamiokande Collaboration), Phys. Rev. Lett.,
	\textbf{81} (1998) 1562. R. Wendell \textit{et al.} 
	(Super-Kamiokande 		Collaboration),
	Phys. Rev., \textbf{D81} (2010) 092004.
	
  	%\bibitem[2]{SM}
 	%D.~Colladay and V.A.~Kostelecky,
  	%``CPT violation and the standard model,''
  	%Phys. Rev., {\bf D55} (1997) 6760
  	%[arXiv:hep-ph/9703464].
  	
  	
  	\bibitem[2]{SM}
  	W.~Buchmuller and C.~Ludeling, (2006)
  	%``Field Theory and Standard Model,''
  	[hep-ph/0609174].
  	
	\bibitem[3]{Kostelecky_cpt}
  V.~A.~Kostelecky and M.~Mewes,
  %``Lorentz and CPT violation in neutrinos,''
  Phys.\ Rev.\ {\bf D69} (2004) 016005
  [hep-ph/0309025].

  	%\bibitem[3]{CPT}
  	%V.A. Kostelecky and R. Lehnert, Phys. Rev.,  \textbf{D63}  (2001) 065008 [arXiv:0012060].
  	
	\bibitem[4]{g-factor_exp}
  	H.~Dehmelt,
  	%``Experiments with an isolated subatomic particle at rest,''
  	Rev.\ Mod.\ Phys.,\  {\bf 62} (1990) 525.
  	
  	\bibitem[5]{proton_antiproton}
  	G. Gabrielse, Int. J. Mass Spectr., {\bf251} (2006) 273-280.

	\bibitem[6]{blaum}
	K. Blaum,  Yu.N. Novikov and G. Werth, Contemp.Phys., {\bf51} (2010) 149-175 [arXiv:0909.1095].

	\bibitem[7]{Greenberg}
  	O.W.~Greenberg,
  	%``CPT violation implies violation of Lorentz invariance,''
  	Phys.\ Rev.\ Lett.,\  {\bf 89} (2002) 231602
  	[hep-ph/0201258].

	
	\bibitem[8]{wp_st_neut}
  K.N.~Abazajian, M.A.~Acero, S.K.~Agarwalla, A.A.~Aguilar-Arevalo, C.H.~Albright, S.~Antusch, C.A.~Arguelles and A.B.~Balantekin {\it et al.},
  %``Light Sterile Neutrinos: A White Paper,''
  [arXiv:1204.5379 [hep-ph]].
	
	\bibitem[9]{st_anom}
  	G.~Mention, M.~Fechner, T.~Lasserre, T.A.~Mueller, D.~Lhuillier, M.~Cribier 		and A.~Letourneau,
  	%``The Reactor Antineutrino Anomaly,''
  	Phys.\ Rev.,\  {\bf D83} (2011) 073006
  	[arXiv:1101.2755 [hep-ex]].
	
  	\bibitem[10]{lsnd_miniboone}
	J.M.~Conrad, W.C.~Louis and M.H.~Shaevitz,
  	%``The LSND and MiniBooNE Oscillation Searches at High $\Delta m^2$,''
  	Ann.\ Rev.\ Nucl.\ Part.\ Sci.\  {\bf 63} (2013) 45
  	[arXiv:1306.6494 [hep-ex]].
	
	
  
  	
	\bibitem[11]{3+1}
C.~Giunti and M.~Laveder,
  %``3+1 and 3+2 Sterile Neutrino Fits,''
  Phys.\ Rev.\ {\bf D84} (2011) 073008
  [arXiv:1107.1452 [hep-ph]].
	
	\bibitem[12]{Kopp:2013vaa}
  	J.~Kopp, P.A.N.~Machado, M.~Maltoni and T.~Schwetz,
  	%``Sterile Neutrino Oscillations: The Global Picture,''
  	JHEP {\bf 1305} (2013) 050
  	[arXiv:1303.3011 [hep-ph]].
	
	\bibitem[13]{sox}
  G.~Bellini {\it et al.}  [Borexino Collaboration],
  %``SOX: Short distance neutrino Oscillations with BoreXino,''
  JHEP {\bf 1308} (2013) 038
  [arXiv:1304.7721 [physics.ins-det]].
	
	\bibitem[14]{neu_oscil}
	Yu.N. Novikov, T. Enqvist, A.N. Erykalov, F. v.Feilitzsch, J. Hissa, K. Loo, D. A. Nesterenko, L. Oberauer \it et al., \rm (2011) [arXiv:1110.2983].
	
	\bibitem[15]{Giunti_cpt}
  	C.~Giunti and M.~Laveder,
  %``Hint of CPT Violation in Short-Baseline Electron Neutrino Disappearance,''
  	Phys.\ Rev.\ {\bf D82} (2010) 113009
  	[arXiv:1008.4750 [hep-ph]].

 	\bibitem[16]{gallex}
	W. Hampel \textit{et al.} (GALLEX),
	%"Final results of the Cr-51 neutrino source experiments in GALLEX",
	Phys. Lett., \textbf{B420} (1998) 114.
	
	\bibitem[17]{gall_51cr}
	R. Bernabei \textit{et al.} (GALLEX),
	%"Implications of the GALLEX results after the chromium source experiment",
	Nucl. Phys. Proc. Suppl., \textbf{48} (1996) 304.

	\bibitem[18]{gno}
	M. Altmann \textit{et al.} (GNO),
	%"Complete results for five years of GNO solar neutrino observations",
	Phys. Lett., \textbf{B616} (2005) 174
	[hep-ex/0504037].

	\bibitem[19]{Pr_data}
	A.A. Sonzogni. Nuclear Data Sheets, {\bf93}  (2001) 599.	 	

  	\bibitem[20]{Ce_source}
  M.~Cribier, M.~Fechner, T.~Lasserre, A.~Letourneau, D.~Lhuillier, G.~Mention, D.~Franco and V.~Kornoukhov {\it et al.},
  %``A proposed search for a fourth neutrino with a PBq antineutrino source,''
  Phys.\ Rev.\ Lett.\  {\bf 107} (2011) 201801
  [arXiv:1107.2335 [hep-ex]].
  	
	\bibitem[21]{spect}
	M. Durero, Master of Science Thesis, Stockholm, Sweden (2013).
	
	\bibitem[22]{51cr}
	H. Xiaolong. Nuclear Data Sheets, {\bf107}  (2006) 2131.
	
	\bibitem[23]{Ce_new_paper}
J.~Gaffiot, T.~Lasserre, G.~Mention, M.~Vivier, M.~Cribier, M.~Durero, V.~Fischer and A.~Letourneau {\it et al.},
  %``Experimental Parameters for a Cerium 144 Based Intense Electron Antineutrino Generator Experiment at Very Short Baselines,''
  (2014) [arXiv:1411.6694 [physics.ins-det]].
  	
	\bibitem[24]{elastic_scatter_form}
  	J.N.~Bahcall, M.~Kamionkowski and A.~Sirlin,
  	%``Solar neutrinos: Radiative corrections in neutrino - electron scattering experiments,''
  	Phys.\ Rev.,\  {\bf D51} (1995) 6146
  	[arXiv:astro-ph/9502003].
 
	%\bibitem[21]{elastic_cross}
	%R.C. Allen \textit{et al.}, Phys. Rev.,  \textbf{D47}  (1993) 11.

	\bibitem[25]{IBD_cross}
	P. Vogel and J.F. Beacom. Phys. Rev.,  \textbf{D60}  (1999) 053003 [arXiv:hep-ph/9903554].
	
 	\bibitem[26]{juno}
 Y.F.~Li,
  %``Overview of the Jiangmen Underground Neutrino Observatory (JUNO),''
  Int.\ J.\ Mod.\ Phys.\ Conf.\ Ser.\  {\bf 31} (2014) 1460300
  [arXiv:1402.6143 [physics.ins-det]].

	\bibitem[27]{Wurm}
	M. Wurm \it et al. \rm [LENA Collaboration], Astropart. Phys., {\bf35} (2012) 685 [arXiv:1104.5620].
	
 	\bibitem[28]{oscil_source}
	J.D. Vergados and Yu.N. Novikov. Nucl. Phys., \bf B839 \rm (2010) 1 [arXiv:1006.3862].
	

	
  	%\bibitem[25]{borexino}
  	%G. Bellini \textit{et al.} (Borexino Collaboration). Phys. Rev. D \textbf{89}, 112007 (2014).



  	%\bibitem[26]{w_phd}
  	%M. Wurm, PhDThesis, Technische Universit\"{a}t M\"{u}nchen, Germany (2009).

\end{thebibliography}



\end{document}